\documentclass{article}
\begin{document}

\begin{center}
\centerline{\large \bf Comment on "Resonant multi-photon }
\centerline{\large \bf IR dissociation spectroscopy of a trapped }
\centerline{\large \bf and sympathetically cooled biomolecular 
ion species"}
\end{center}

\vspace{3 pt}
\centerline{\sl V.A. Kuz'menko\footnote{Electronic 
address: kuzmenko@triniti.ru}}

\vspace{5 pt}

\centerline{\small \it Troitsk Institute for Innovation and Fusion 
Research,}
\centerline{\small \it Troitsk, Moscow region, 142190, Russian 
Federation.}

\vspace{5 pt}

\begin{abstract}

In the e-print arXiv:1110.2774 the authors study the infrared 
multiple-photon induced dissociation (IRMPD) of cooled 
polyatomic ion species. Here we propose a physical 
explanation of the obtained experimental results 
and its simple experimental test.

\end{abstract}

\vspace{12 pt}

In the works [1, 2] the vibrational spectroscopy of cooled 
polyatomic ion (protonated dipeptide tryptophane-alanine) 
in the region 2.74 $\mu m$ was studied by two similar, 
but little different methods. In the first case [1] the 
two-step IR-UV photodissociation was used. In other case 
[2] the one-step IRMPD was studied and there was discovered 
that the cross-section for IRMPD is approximately two orders 
smaller and spectral width is considerably larger than in 
the first case.

The goal of this comment is to give a possible correct 
physical explanation of the discovered difference. The 
main reason for these differences is the use of the infrared 
radiation of different spectral width. In the first case the 
spectral linewidth of the pulsed IR laser radiation was near 
$\sim 1 cm^{-1}$ and this radiation can efficiently interact 
with great number of narrow rotational lines of cold molecules. 
And there was sufficiently to absorb one photon of IR 
radiation for the following dissociation after the 
absorption of UV photon.

In the second case the continuous wave IR radiation with 
narrow spectral linewidth ($<100$ kHz) was used. This 
radiation efficiently interacts only with small number 
of narrow rotational lines and easily saturates it. However, 
for the dissociation the molecule should absorb a number of 
IR photons. These process of collisionless multiple photon 
absorption proceeds through the so-called wide component of 
lines or line wings. 

The existence of this wide component of lines was discovered 
in the middle infrared region ($\sim 10 \mu m $) in works 
[3-5]. A physical nature of this wide component is not 
sufficiently clear till now. It has Lorentzian shape with usual 
width of several $cm^{-1}$. Its relative integral intensity 
depends from the number of atoms, branching degree of the molecule 
and is changed from $\sim 0.6\%$  for  $SF_6$ and $SiF_4$  to 
$\sim 90\%$ for  $(CF_3)_2CO$ molecules. 

So, the difference in experimental results is, probably, 
determined by the fact that in the first case [1] the authors 
record an integral spectrum in contrast to the second case [2], 
where the authors deal mainly with the wide component, which 
can contain only small part of integral spectrum. There is 
simple experimental test for the proposed physical explanation. 
The experiments of two-step IR-UV dissociation may be 
compared for two variants: with broadband IR laser radiation 
(as in [1]) and with the same energy but with narrow 
linewidth IR laser radiation ($<100$ kHz). 

Unfortunately, nobody studies today the wide component 
of lines, although, this is interesting and important 
physical object:

1) the wide component of lines exists, probably, in spectra of all 
polyatomic molecules,

2) it in a substantial degree affects the climate of the Earth [6],

3) in the experiments with this object [4] there was obtain the 
first direct experimental proof of the important property of 
quantum physics - time reversal noninvariance [7].

A molecular beam with cryogenic bolometer [4] is a 
most suitable developed method for experimental study 
of the wide component of lines. However, the used in 
[1, 2] method of IR-UV and IRMPD of trapped cold 
polyatomic ions can have also some advantages: 
extremely low temperature ($<1^{0}$ K) and very long 
collisionless trapping time, which allows to use 
low intensity continuous wave IR laser radiation.

\end{document}